\documentclass[%
 reprint,
 amsmath,amssymb,
 aps,
]{revtex4-2}

\usepackage{graphicx}
\usepackage{dcolumn}
\usepackage{bm}
\usepackage{amsfonts}
\usepackage{amsmath}

\usepackage{subfigure}
\usepackage{hyperref}
\usepackage{float}

\begin{document}

\preprint{APS/123-QED}

\title{Commutation Relations in Adiabatic Elimination}

\author{Hong Xie$^{1*}$}
\author{Le-Wei He$^{1}$}
\author{Xiu-Min Lin$^{2,3,4 \dag}$}

\affiliation{$^{1}$ Department of Mathematics and Physics, Fujian Jiangxia University, Fuzhou 350108, China}
\affiliation{$^{2}$ Fujian Provincial Key Laboratory of Quantum Manipulation and New Energy Materials, College of Physics and Energy, Fujian Normal University, Fuzhou 350117, China}
\affiliation{$^{3}$  Fujian Provincial Engineering Technology Research Center of Solar Energy Conversion and Energy Storage, Fuzhou,350117, China.}
\affiliation{$^{4}$  Fujian Provincial Collaborative Innovation Center for Advanced High-Field Superconducting Materials and Engineering, Fuzhou 350117, China}

\begin{abstract}
The method of adiabatic elimination has been widely adopted in quantum optics in the past several decades. In the study of cavity-based light-matter interactions, the bad-cavity limit is often encountered, where the damping rate of the cavity is much larger than the interaction strength. The fast-damped cavity will quickly relax to a quasi-stationary state, and one can eliminate the cavity from the equation of motion by setting its time derivative to zero. Elimination of the cavity in the bad-cavity limit can reduce the dimensionality of the equations of motion of the system. However, we find that the adiabatic elimination procedure performed in the quantum Langevin equation leads to an incorrect commutation relation, which was rarely discussed in the former studies, as far as we know. Here, we show the incorrect commutation relation arises from the fact that the high frequency of the vacuum noise should be cut off to perform adiabatic elimination, but the noise with high frequency cutoff is still treated as white noise with infinite bandwidth and delta commutation relation. We also study the correlation function and show that the high frequency part of noise contributes very little when averaged over the bath. Therefore, the adiabatic elimination method can reduce the complexity of the calculations while maintaining physical reliability.

\end{abstract}
\maketitle



\section{introduction}
There are a number of situations in interaction systems where some variables vary much more rapidly on time scales than others. In an open system with dissipations, if the dynamics of one variable damps much faster than the rate at which another variable evolves, the fast variable can relax to a quasi-stationary state, in which the values of the fast variable follow those of the slow variables. It is said that the fast variable is ``slaved" to the slow variables \cite{Haken2004}, and the fast variable can be replaced by its steady-state values, being functions of the slow variables, leaving the equation of motion for the slow variables alone. This approximation method is the so-called adiabatic elimination. Eliminating the fast variables from the equation of motion can reduce the dimensionality of a problem, sometimes to the point where analytical solutions become accessible.

The method of adiabatic elimination was originally developed and deeply studied in statistical mechanics \cite{Haken_1977,haake1982systematic,sancho1982adiabatic,gardiner1984adiabatic,lugiato1984adiabatic,gough2007singular}. In quantum optic systems, adiabatic elimination was early used in the development
of a theory of the maser and laser that includes quantum noise \cite{lax1966quantum}, and becomes a common technique in quantum optics in the past several decades \cite{wiseman2009quantum,gardiner2004quantum}. Especially, in the studies of light-matter interactions in the bad-cavity limit, adiabatic elimination of the fast-damped cavity mode has been utilized in various systems, for example, atomic ensemble \cite{duan2001long,RevModPhys.83.33}, cavity QED \cite{PhysRevA.66.022314,doherty1998motional,duan2004scalable}, and cavity optomechanics \cite{PhysRevA.84.052327,PhysRevLett.112.143602}. 

Eliminating the cavity mode in the quantum Langevin equation is usually carried out by setting the time derivative of the cavity mode to zero, then the algebraic equation of the cavity mode includes vacuum noise. The delta-commutated vacuum noise will lead to a divergent commutator of the cavity operator, which is inconsistent with the canonical commutation relations. Although the elimination of the cavity mode in the bad-cavity limit is considered to be a good approximation, to our knowledge, the incorrect commutation relation in adiabatic elimination has rarely been discussed in previous studies. The commutation relations are a fundamental issue in quantum mechanics, and resolving this ambiguity arising in adiabatic elimination is important.

Here, we study the adiabatic elimination method in the quantum Langevin equation and show that the elimination of the cavity mode requires the treatment of the vacuum noise as a slowly varying variable. This is incompatible with the fact that vacuum noise is a delta-correlated quantum white noise. In the frequency domain, this means the cavity has a finite bandwidth, which is not strictly possible to be ``slaved" to the vacuum noise that corresponds to the infinite bandwidth. To eliminate the cavity mode, the high frequency part of the vacuum noise should be cut off. The divergent commutation relation of the cavity operator appears when the high frequency cutoff noise is still treated effectively as white noise.  Physically, only the frequencies that are close to the resonant frequency of the system are important; the high frequency part of the noise has little effect on the measurable quantities. We study the correlation function of the cavity mode and find that the high frequency of the vacuum noise plays a negligible role. Using the delta function expansion, we also show that the correlation function obtained by adiabatic elimination is the leading order term of the rigorous result.


The paper is arranged as follows. Eliminating the cavity mode in the interaction system and the incorrect commutation relations are discussed in Sec.~\ref{2}. The origin of incorrect commutation is analyzed in Sec.~\ref{3}. The high-frequency cutoff of vacuum noise is introduced in Sec.~\ref{4}. Section \ref{5} studies the influence of the high-frequency part of the noise on the correlation function, and Section \ref{6} gives the conclusions.

%


\section{Adiabatic elimination of the cavity mode}\label{2}
We consider a widely studied interaction Hamiltonian in quantum optics
 \begin{eqnarray}\label{interaction}
H_\text{I} =  \hbar G(ab^\dag+a^\dag b),
\end{eqnarray}
where $a$ is the annihilation operator of the cavity mode, $G$ denotes the coupling strength, and $b$ is a general representation of the operators for different systems. The operator $b$ could be lowering operator $\sigma_{-}$ of the two-level atom, then the Hamiltonian $H_\text{I}$ represents the interaction part of the well-known Jaynes-Cummings model \cite{gerry2023introductory}. The operator $b$ could also be the lowering operator of the collective atom mode in atomic ensembles \cite{duan2001long} or the annihilation operator of mechanical mode in cavity optomechanical systems \cite{RevModPhys.86.1391}.

The corresponding quantum Langevin equations for the Hamiltonian $H_\text{I}$ are given by
 \begin{subequations}\label{Langevin}
 \begin{align}
\dfrac{da}{dt}=-iGb-\dfrac{\kappa}{2}a+\sqrt{\kappa}a_\text{in},\\
 \dfrac{db}{dt}=-iGa-\dfrac{\gamma}{2}b+\sqrt{\gamma}b_\text{in},\label{Langevin2}
\end{align} 
\end{subequations}
where $\kappa$ and $\gamma$ are the damping rates of cavity and system $b$, respectively. $a_\text{in}$ is the vacuum input noise for the cavity, and $b_\text{in}$ denotes the stochastic force added to system $b$. 

In the case of $\gamma \ll \kappa$ and bad-cavity limit $G\ll \kappa$, the cavity is damped at a rate much faster than the rate at which the system $b$ changes the cavity state. Then the cavity will reach a steady state, which will adjust to the changes in system $b$. We can eliminate the cavity by setting $da/dt=0$ and obtain
 \begin{eqnarray}\label{elimination}
a=-i\frac{2G}{\kappa}b+\frac{2}{\sqrt{\kappa}}a_\text{in},
\end{eqnarray}
which can be substituted into Eq.~(\ref{Langevin2}), leaving the equation of motion for system $b$ alone. The adiabatic elimination method in the quantum Langevin equation is simply to set the time derivative of the cavity to zero, and it can reduce the dimension of the equation of motion of the system. This method has been widely used in various light-matter interaction systems. In the adiabatic elimination procedure, the cavity mode $a$ follows the evolution of $b$, which is a good approximation since $G \ll \kappa$. However, one should be careful when dealing with vacuum noise. The vacuum noise has delta-correlated commutation relations 
 \begin{eqnarray}\label{delta-correlated}
[a_\text{in}(t),a_\text{in}^\dag(t')]=\delta(t-t').
\end{eqnarray}
This leads to a divergent commutator of the equal-time operator $a$ in Eq.~(\ref{elimination}),
 \begin{eqnarray}\label{commutation_noise}
[a(t),a^\dag(t)] \rightarrow \infty,
\end{eqnarray}
which is inconsistent with the canonical commutation relations $[a(t),a^\dag(t)]=1$. Since the commutation relation plays a fundamental role in quantum mechanics, it is important to study the approximation that is made in adiabatic elimination and check whether it is reasonable.

\section{Commutation relations in adiabatic elimination}\label{3}
The divergent commutation relations result from the input vacuum noise $a_\text{in}(t)$, which should be treated carefully because of its delta-correlated characteristics. In the following, for simplicity, we consider only a bare cavity coupled to its heat bath and follow the treatment of the references \cite{collett1984squeezing,gardiner2004quantum,walls2008input} for the definition of the vacuum noise. The system Hamiltonian of a bare cavity coupled to the bath is described by
\begin{eqnarray}\label{Hamiltonian}
H=H_c+H_\text{bath}+H_\text{int}
\end{eqnarray}
with the Hamiltonian of the bare cavity
\begin{eqnarray}\label{Hamiltonian_c}
H_c=\hbar\omega_c a^\dag a,
\end{eqnarray}
the free Hamiltonian of the bath
\begin{eqnarray}\label{Hamiltonian_c}
H_\text{bath}=\hbar\int_{-\infty}^{\infty}d\omega\omega b^\dag(\omega) b(\omega),
\end{eqnarray}
and the interaction Hamiltonian
\begin{eqnarray}\label{Hamiltonian_c}
H_\text{int}=i\hbar\int_{-\infty}^{\infty}d\omega g(\omega)[ b^\dag(\omega)a - b(\omega)a^\dag],
\end{eqnarray}
where $b(\omega)$ is the annihilation operator of the bath with commutation relations $[b(\omega), b^\dag(\omega')]=\delta(\omega-\omega')$, and $g(\omega)$ is coupling constant. In fact, the range of the frequency $\omega$ integration should be $(0,\infty)$. However, the terms far from the resonance of the system are negligible under the rotating wave approximation. This allows one to extend the range of integration to $(-\infty,\infty)$, which is essential for the vacuum noise with delta commutation relations.

The Heisenberg equations of motion for the cavity and bath are
\begin{subequations}\label{Heisenberg}
 \begin{align}
\dfrac{da}{dt}=-i\omega_c a-\int_{-\infty}^{\infty}d\omega g(\omega)b(\omega),\\
\dfrac{db(\omega)}{dt}=-i\omega b(\omega)+g(\omega)a. \label{Heisenberg1}
 \end{align}
\end{subequations}
In terms of the solution of Eq.~(\ref{Heisenberg1}), the cavity operator obeys the equation
\begin{eqnarray}\label{a}
\dfrac{da}{dt}=&&-i\omega_c a-\int_{-\infty}^{\infty}d\omega g(\omega)e^{-i\omega t}b_0(\omega)
 \nonumber \\ &&-\int_{-\infty}^{\infty}d\omega g^2(\omega)\int_{0}^{t}dt' e^{-i\omega (t-t')}a(t'),
\end{eqnarray}
where $b_0(\omega)$ is the value of $b(\omega)$ at $t=0$. By introducing the first Markov approximation that the coupling $g(\omega)$ is assumed to be independent of frequency, then we set $g(\omega)=\sqrt{\kappa/2\pi}$. Interchanging the order of time and frequency integration in the last term of Eq.~(\ref{a}) and making a transformation $a\to ae^{-i\omega_ct}$, the equation becomes
\begin{eqnarray}\label{a_Langevin}
\dfrac{da}{dt}=-\frac{\kappa}{2}a+\sqrt{\kappa}a_\text{in},
\end{eqnarray}
with the input vacuum noise in the rotating frame of $\omega_c$ is defined as
\begin{eqnarray}\label{ain_123}
a_{\text{in}}(t) =\frac{-1}{\sqrt{2\pi}} \int_{-\infty}^{\infty}{d\omega e^{i(\omega_c-\omega)t} b_0(\omega)}.
\end{eqnarray}
As mentioned above, the integration limits $(-\infty,\infty)$ is necessary to ensure $a_{\text{in}}(t)$ being a white noise with delta commutation relations. Using $[b_0(\omega), b_0^\dag(\omega')]=\delta(\omega-\omega')$ and $\int_{-\infty}^{\infty} d\omega e^{-i\omega (t-t')} = 2\pi \delta(t-t')$, one can obtain the commutation relation in Eq.~(\ref{delta-correlated}). The delta correlation function $\langle a_{\text{in}}^\dag(t) a_{\text{in}}(t')\rangle = \bar{n}_\text{th}\delta (t-t')$ can also be obtained using $\langle b_0^\dag(\omega) b_0(\omega')\rangle = \bar{n}_\text{th}\delta (\omega-\omega')$, where $\bar{n}_\text{th}$ is the average occupation number of the bath.


Based on Eqs.~(\ref{a_Langevin}) and (\ref{ain_123}), we now discuss the commutation relations in adiabatic elimination.
When adiabatic elimination is performed in Eq.~(\ref{a_Langevin}), we set $da/dt=0$ and obtain 
\begin{eqnarray}\label{d_adiabatic}
a(t)=\frac{2}{\sqrt{\kappa}}a_{\text{in}}(t),
\end{eqnarray}
which results in a divergent commutation relation.
If Eq.~(\ref{a_Langevin}) is integrated without initial condition, we obtain
\begin{eqnarray}\label{a1full}
a(t) = \sqrt{\kappa}\int_{-\infty}^{t}{d\tau e^{-\kappa(t-\tau)/2} a_{\text{in}}(\tau)}.
\end{eqnarray}
Using the delta commutation relations of $a_{\text{in}}(t)$, we obtain the correct commutation relations $[a(t),a^\dag(t)]=1$.  The result of making adiabatic elimination Eq.~(\ref{d_adiabatic}) can also be obtained by treating $a_{\text{in}}(\tau)$ as a slowly varying function and replacing $a_{\text{in}}(\tau)$ by $a_{\text{in}}(t)$ outside the integral
\begin{eqnarray}
a(t) = \sqrt{\kappa}a_{\text{in}}(t)}\int_{-\infty}^{t}{d\tau e^{-\kappa(t-\tau)/2} = \frac{2}{\sqrt{\kappa}}a_{\text{in}}(t).
\end{eqnarray}
This means that making adiabatic elimination corresponds to treating $a_{\text{in}}(t)$ as a slowly varying function, i.e., the time scales of interest in $a_{\text{in}}(t)$ is much larger than $\kappa^{-1}$. However, the vacuum noise is delta-correlated, $a_{\text{in}}(t)$ and $a_{\text{in}}(t-\kappa^{-1})$ are not only different, they are uncorrelated. In other words, the vacuum noise $a_{\text{in}}$ has infinite bandwidth, so it is not strictly possible to slave $a$ that only responds to a finite bandwidth $\kappa$, to the vacuum fluctuation \cite{wiseman2009quantum}. Therefore, the high frequency of the vacuum noise should be cut off in order to perform the adiabatic elimination, but the high frequency cutoff noise is still effectively treated as white noise with delta commutation relations. This leads to the divergent commutation relations of the cavity operator $a$.

\begin{figure}[t]
 {\includegraphics[width=0.45\textwidth]{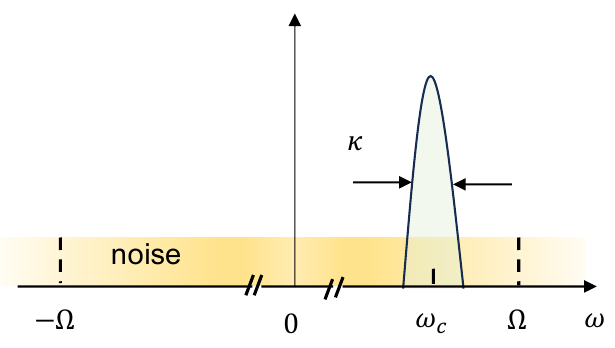}}
\caption{Schematic of the spectrums of vacuum noise and cavity mode. The infinite bandwidth of vacuum noise is cut off, and the integration range of the noise is set to $(-\Omega,\Omega)$.}
\label{spectrum}
\end{figure} 

\section{High frequency cutoff}\label{4}

\begin{figure}[t]
 {\includegraphics[width=0.5\textwidth]{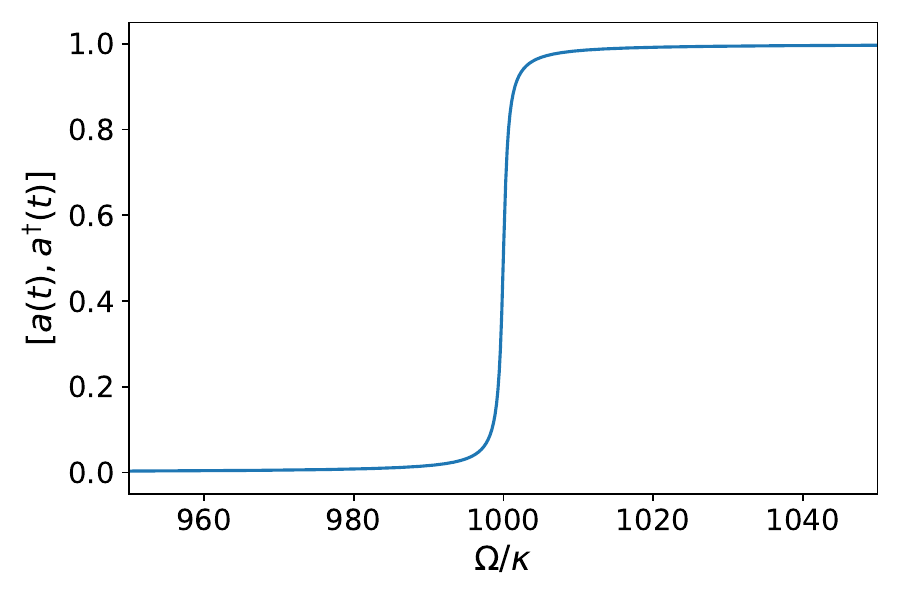}}
\caption{The commutator Eq.~(\ref{a_c12}) with high frequency cutoff  is plotted as a function of $\Omega/\kappa$. The quality factor of the cavity is chosen as $\omega_c/\kappa=10^3$.}
\label{commutator}
\end{figure}

To study the influences of high frequency of the noise on commutation relations of the cavity operator, we set a high frequency cutoff to the noise and restrict the range of integration from $(-\infty,\infty)$ to $(-\Omega,\Omega)$, as shown in Fig.~\ref{spectrum}. Then the vacuum noise in Eq.~(\ref{ain_123}) is redefined as
\begin{eqnarray}\label{ain_cutoff}
a_{\text{in}}(t) =\frac{-1}{\sqrt{2\pi}} \int_{-\Omega}^{\Omega}{d\omega e^{i(\omega_c-\omega)t} b_0(\omega)},
\end{eqnarray}
and Eq. (\ref{a_Langevin}) is integrated as
\begin{eqnarray}\label{a_12}
a(t) = -&&\sqrt{\frac{\kappa}{2\pi}} e^{-\kappa t/2} \int_{-\infty}^{t}{d\tau e^{\kappa  \tau/2 }}\\ \nonumber
&&\times \int_{-\Omega}^{\Omega}d\omega e^{i(\omega_c-\omega)\tau}b_0(\omega).
\end{eqnarray}
Interchanging the order of time and frequency integration, the commutator of the cavity operator can be calculated as
\begin{eqnarray}\label{a_c}
[a(t), a^{\dag}(t)] = {\frac{\kappa}{2\pi}} \int_{-\Omega}^{\Omega}d\omega \frac{1}{(\omega-\omega_c)^2+\kappa^2/4}.
\end{eqnarray}
The solution of Eq.~(\ref{a_c}) is given by
\begin{eqnarray}\label{a_c12}
[a(t), a^{\dag}(t)] = \frac{1}{\pi}[&&\arctan\frac{2}{\kappa}(\Omega-\omega_c)\\ \nonumber
&&+\arctan\frac{2}{\kappa}(\Omega+\omega_c)].
\end{eqnarray}
Based on the properties of the inverse trigonometric function arctan(x), it is known that the high frequency part of vacuum noise contributes little to the commutator. The commutator $[a(t), a^{\dag}(t)]$ as a function of $\Omega/\kappa$ is plotted in Fig.~\ref{commutator}, where a bad cavity with quality factor $\omega_c/\kappa=10^3$ is considered. The commutator quickly approaches $1$ at $\Omega/\kappa = 10^3$, and it almost equals to $1$ when $\Omega/\kappa > \omega_c/\kappa$. The exact communication relation $[a_1(t), a^{\dag}_1(t)]=1$ is obtained in the limit of $\Omega \rightarrow \infty$, as expected.

When the adiabatic elimination is carried out to solve the Langevin equation, it means that the high frequency part of vacuum noise should be omitted in the elimination procedure, but the noise is effectively white. This is equivalent to making an approximation in Eq.~(\ref{a_c}) by neglecting the term $(\omega-\omega_c)^2$ (only the terms near resonance are considered), and keeping the range of the integration $(-\infty,\infty)$ which is essential for the definition of white noise, namely, $[a(t), a^{\dag}(t)] = {\frac{\kappa}{2\pi}} \int_{-\infty}^{\infty}d\omega \frac{4}{\kappa^2}$.
This gives rise to a divergent commutator, which is similar to the result obtained from the adiabatically eliminated expression $a= (2/\sqrt{\kappa})a_{\text{in}}$. 

We can now conclude that the contribution of the high frequency part of the noise on the commutator is negligibly small under normal circumstances, but when adiabatic elimination is performed, the elimination over the infinite bandwidth of the vacuum noise leads to a divergent commutator.

\section{Correlation function}\label{5}
Although an incorrect communication relation appears when the high frequency is omitted in the adiabatic elimination procedure, it should be noted that the measurable quantities which average over 
the bath are usually affected a little by the high frequency part of the noise. We study the correlation function of the cavity for example. 

\subsection{Correlation function without high frequency cutoff}

Based on the result of directly integrating the Langevin equation Eq.~(\ref{a1full}), the correlation function of the cavity is expressed as
\begin{eqnarray}\label{a_correlation}
\langle a^{\dag}(t)a(t')\rangle = && {\kappa}\int_{-\infty}^{t}\int_{-\infty}^{t'}d\tau d\tau' \Big[e^{-\kappa(t-\tau)/2} e^{-\kappa(t'-\tau')/2}  \nonumber \\&&
\times \langle a^{\dag}_{\text{in}}(\tau) a_{\text{in}}(\tau')\rangle \Big].
\end{eqnarray}
For the white noise Eq.~{(\ref{ain_123})} with infinite integration $(-\infty,\infty)$, it is delta-correlated $\langle a_{\text{in}}^\dag(t) a_{\text{in}}(t')\rangle = \bar{n}_\text{th}\delta (t-t')$, and the correlation function can be integrated as
\begin{eqnarray}\label{a_correlation1}
\left< a^{\dag}(t)a(t')\right> = \bar{n}_{\text{th}}e^{-\kappa|t-t'|/2},
\end{eqnarray}
and the equal-time correlation function is $\langle a^{\dag}(t)a(t)\rangle = \bar{n}_{\text{th}}$.

\subsection{Correlation function with high frequency cutoff}

\begin{figure}[!t]
 {\includegraphics[width=0.5\textwidth]{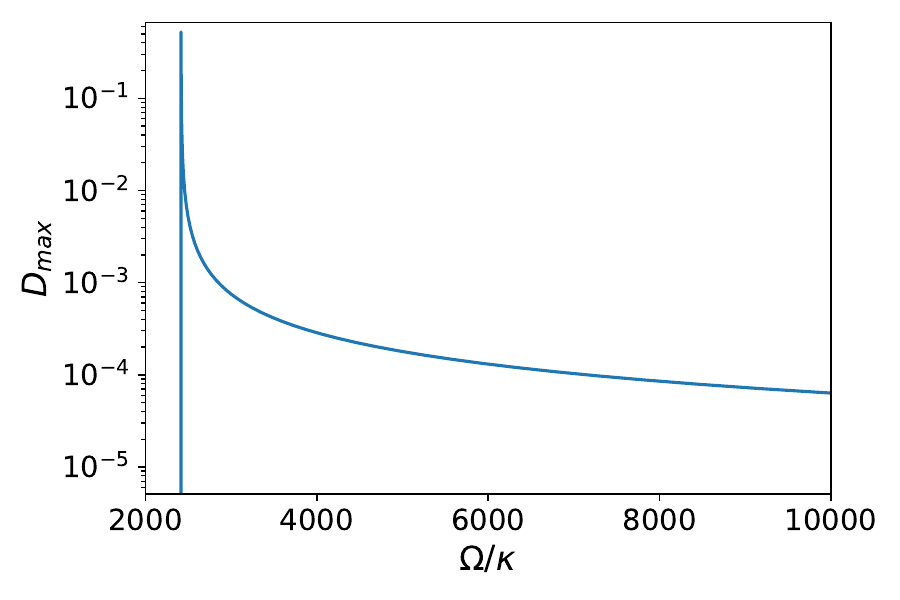}}
\caption{The maximal value of difference $D$ as a function of $\Omega/\kappa$. The quality factor of the cavity is $\omega_c/\kappa = 10^3$.}
\label{D_max}
\end{figure}

When the high frequency cutoff is introduced to the bath, the vacuum noise is given by Eq.~(\ref{ain_cutoff}), using the result of Eq.~(\ref{a_12}), the correlation function is expressed as
\begin{eqnarray}\label{a_correlation2}
 \langle a^{\dag}(t)a(t')\rangle_\text{cut} = && \frac{\kappa}{2\pi}e^{-\frac{\kappa}{2}(t+t')}
 \int_{-\Omega}^{\Omega}d\omega  \int_{-\Omega}^{\Omega}d\omega' 
 \int_{-\infty}^{t}d\tau \nonumber \\&&
 \times \int_{-\infty}^{t'} d\tau' 
\{ e^{[-i(\omega_c-\omega)+\frac{\kappa}{2}]\tau} 
 \nonumber \\&&
\times e^{[i(\omega_c-\omega')+\frac{\kappa}{2}]\tau'}\langle b^{\dag}_{0}(\omega) b_{0}(\omega')\rangle\}.
\end{eqnarray}
Based on the delta-correlator $\langle b^{\dag}_{0}(\omega) b_{0}(\omega')\rangle=\bar{n}_{\text{th}}\delta(\omega-\omega')$ of the thermal bath, the correlation function becomes
\begin{eqnarray}\label{a_correlation2}
\langle a^{\dag}(t)a(t')\rangle_\text{cut} = \frac{\kappa}{2\pi}\bar{n}_{\text{th}}f(t-t'), 
\end{eqnarray}
where
\begin{eqnarray}\label{ft3}
f(t-t')= 
\int_{-\Omega}^{\Omega}d\omega \frac{e^{i(\omega-\omega_c)(t-t')}}{(\omega-\omega_c)^2+\kappa^2/4}  .
\end{eqnarray}
The function $f(t-t')$ can be calculated using the residue theorem in complex analysis \cite{stein2010complex}. Here we only give the results, the detailed calculation is presented in the Appendix. For sufficient large $\Omega$, the function is given by 
\begin{eqnarray}\label{ft2}
f(t-t') = \frac{2\pi}{\kappa}e^{-\frac{\kappa}{2}|t-t'|}+ S,
\end{eqnarray}
where $S$ denotes a small term that is bounded by
\begin{eqnarray}\label{small}
S \leq  \frac{\pi\Omega}{\Omega^2-2\omega_c\Omega-\omega^2_c-\kappa^2/4}.
\end{eqnarray}
Then the difference between correlation function without and with high frequency cutoff is
\begin{eqnarray}\label{a_correlation2}
D && =  \langle a^{\dag}(t)a(t')\rangle_\text{cut} - \langle a^{\dag}(t)a(t')\rangle 
\nonumber \\ && \leq  \frac{2\Omega/\kappa}{4(\frac{\Omega}{\kappa})^2-8\frac{\omega_c}{\kappa}\frac{\Omega}{\kappa}-4(\frac{\omega_c}{\kappa})^2-1}.
\end{eqnarray}
In the case of $\Omega/\kappa \gg \omega_c/\kappa$, the difference $D$ is negligibly small, as shown in Fig.~\ref{D_max} for the maximal value of $D$. In the limit $\Omega/\kappa \rightarrow \infty$, the difference $D$ will reduce to zero. This means the high frequency of the noise has little effect on the correlation function when traced over the bath. 

\subsection{Correlation function with adiabatic elimination}
Based on the expression $a= (2/\sqrt{\kappa})a_{\text{in}}$ that is straightly obtained using adiabatic elimination, the  correlation function is 
\begin{eqnarray}\label{corre_el}
\langle a^{\dag}(t)a(t')\rangle_\text{el}  = \frac{4 \bar{n}_{\text{th}} }{\kappa}\delta(t-t').
\end{eqnarray}
At first glance, this result is quite different from the rigorous one in Eq.~(\ref{a_correlation1}). However, the exponential function can be expanded in terms of the delta function \cite{lindell1993delta} 
 \begin{eqnarray}\label{delta}
e^{-\frac{\kappa}{2}|t-t'|}=2(\frac{2}{\kappa})\delta(t-t')+2(\frac{2}{\kappa})^{2}\frac{d}{dt'}\delta(t-t')+\cdots 
\end{eqnarray}
for $t>t'$. For large values of $\kappa$, one can make an approximation to keep the terms up to order $2/\kappa$. Then the correlation function in Eq.~(\ref{a_correlation1}) can be approximated to that in Eq.~(\ref{corre_el}). In this sense, the correlation function obtained by the adiabatic elimination method is the leading term of the rigorous result in an expansion in powers of $2/\kappa$.

%

\section{Conclusion}\label{6}
In conclusion, we have studied the ambiguity of the commutation relations arising in adiabatic elimination. We show that the high frequency of the vacuum noise should be cut off to obtain a finite bandwidth, which is necessary to perform the adiabatic elimination, but $a_{\text{in}}$ in the elimination result $a= (2/\sqrt{\kappa})a_{\text{in}}$ is still treated as effective white noise with infinite bandwidth. This leads to the divergence of the commutator $[a(t), a^{\dag}(t)]$. Physically, it is argued that only the frequencies close to the resonance of the system are important. The high frequency part of the noise is far from resonant and contributes very little when averaged over the bath, as discussed above for the correlation function. Therefore, the simple elimination method in the quantum Langevin equation by setting $da/dt=0$ can simplify the calculations while maintaining physical reliability.

\section*{Acknowledgments}
 We acknowledge supports from the National Natural Science Foundation of China (Grants No.~12174054, No.~12074067, and No.~12274079), the Natural Science Foundation of Fujian Province of China (Grants No.~2021J011228 and No.~2022J02027).

\appendix*
\section{Calculation of Eq.~(\ref{ft3}) using residue theorem}

\begin{figure}[h]
 {\includegraphics[width=0.45\textwidth]{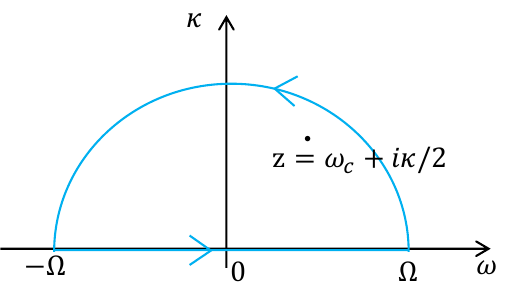}}
\caption{The singularity $z=\omega_c + i\kappa/2$ is in the region bounded by the contour $C$.}
\label{contour}
\end{figure}

 Here, we use the contour integration method to calculate Eq.~(\ref{ft3}). We first consider the case $t-t'>0$, and introduce a contour $C$ that goes along the real line from $-\Omega$ to $\Omega$, then counterclockwise along a semicircle centered at $0$ from $\Omega$ to $-\Omega$, as shown in Fig.~\ref{contour}.  The contour C can be split into a straight part and a curved arc, so that
\begin{eqnarray}
\label{ft} \int_{-\Omega}^{\Omega}d\omega \frac{e^{i(\omega-\omega_c)(t-t')}}{(\omega-\omega_c)^2+\kappa^2/4} \\ \nonumber
= 
 \oint_{C}^{}g(z)dz 
 - \int_{arc}^{}g(z)dz,
\end{eqnarray}
where
\begin{eqnarray}\label{gz}
g(z) = \frac{e^{i(z-\omega_c)(t-t')}}{(z-\omega_c)^2+\kappa^2/4}.
\end{eqnarray}
Since $e^{i(z-\omega_c)(t-t')}$ has no singularities at any point in the complex plane, the function $g(z)$ has singularities only where the denominator ${(z-\omega_c)^2+\kappa^2/4}$ is zero, hence we have two singularities $z=\omega_c \pm i\kappa/2$. Supposed that $\Omega^2 > {\omega_c^2+\kappa^2/4}$, only the point $z=\omega_c + i\kappa/2$ is in the region bounded by the contour $C$. The residue of $g(z)$ at this point is given by 
\begin{eqnarray}\label{res}
\text{Res} \ g(\omega_c + i\kappa/2) = \frac{1}{i \kappa}e^{-\frac{\kappa}{2}(t-t')}.
\end{eqnarray}

According to the residue theorem, we have
\begin{eqnarray}\label{it}
 \oint_{C}^{}g(z)dz &&= 2\pi i \cdot \text{Res} \ g(\omega_c + i\kappa/2) \nonumber \\
&& = \frac{2\pi}{\kappa}e^{-\frac{\kappa}{2}(t-t')}.
\end{eqnarray}
Using some estimations, we have
\begin{eqnarray}\label{arc}
\lvert \int_{arc}^{}g(z)dz \rvert &&= | \int_{arc}^{}\frac{e^{i(z-\omega_c)(t-t')}}{(z-\omega_c)^2+\kappa^2/4}dz |, \nonumber \\
                                               && \le  \int_{arc}^{}\frac{|dz|}{|z|^2-2 \omega_c |z|-\omega^2_c-\kappa^2/4} \nonumber \\
                                               && = \frac{\pi\Omega}{\Omega^2-2\omega_c\Omega-\omega^2_c-\kappa^2/4}.
\end{eqnarray}
In the case of $\Omega^2 \gg \omega_c^2 + \kappa^2/4 $, Eq.~(\ref{arc}) is negligibly small and becomes zero in the limit $\Omega \rightarrow \infty$.

If $t-t'<0$,  a similar argument with an contour $C'$ in the lower half plane shows that
\begin{eqnarray}\label{it}
 \oint_{C'}^{}g(z)dz &&= 2\pi i \cdot \text{Res} \ g(\omega_c - i\kappa/2) \nonumber \\
 &&= \frac{2\pi}{\kappa}e^{\frac{\kappa}{2}(t-t')}.
\end{eqnarray}
We finally have
\begin{eqnarray}\label{final}
 \oint_{C}^{}g(z)dz = \frac{2\pi}{\kappa}e^{-\frac{\kappa}{2} |t-t'|},
\end{eqnarray}
and the small term mentioned in Eq.~(\ref{ft2}) is bound by Eq.~(\ref{arc}).

\bibliography{adiabatic}

\begin{thebibliography}{23}%
\makeatletter
\providecommand \@ifxundefined [1]{%
 \@ifx{#1\undefined}
}%
\providecommand \@ifnum [1]{%
 \ifnum #1\expandafter \@firstoftwo
 \else \expandafter \@secondoftwo
 \fi
}%
\providecommand \@ifx [1]{%
 \ifx #1\expandafter \@firstoftwo
 \else \expandafter \@secondoftwo
 \fi
}%
\providecommand \natexlab [1]{#1}%
\providecommand \enquote  [1]{``#1''}%
\providecommand \bibnamefont  [1]{#1}%
\providecommand \bibfnamefont [1]{#1}%
\providecommand \citenamefont [1]{#1}%
\providecommand \href@noop [0]{\@secondoftwo}%
\providecommand \href [0]{\begingroup \@sanitize@url \@href}%
\providecommand \@href[1]{\@@startlink{#1}\@@href}%
\providecommand \@@href[1]{\endgroup#1\@@endlink}%
\providecommand \@sanitize@url [0]{\catcode `\\12\catcode `\$12\catcode
  `\&12\catcode `\#12\catcode `\^12\catcode `\_12\catcode `\%12\relax}%
\providecommand \@@startlink[1]{}%
\providecommand \@@endlink[0]{}%
\providecommand \url  [0]{\begingroup\@sanitize@url \@url }%
\providecommand \@url [1]{\endgroup\@href {#1}{\urlprefix }}%
\providecommand \urlprefix  [0]{URL }%
\providecommand \Eprint [0]{\href }%
\providecommand \doibase [0]{https://doi.org/}%
\providecommand \selectlanguage [0]{\@gobble}%
\providecommand \bibinfo  [0]{\@secondoftwo}%
\providecommand \bibfield  [0]{\@secondoftwo}%
\providecommand \translation [1]{[#1]}%
\providecommand \BibitemOpen [0]{}%
\providecommand \bibitemStop [0]{}%
\providecommand \bibitemNoStop [0]{.\EOS\space}%
\providecommand \EOS [0]{\spacefactor3000\relax}%
\providecommand \BibitemShut  [1]{\csname bibitem#1\endcsname}%
\let\auto@bib@innerbib\@empty
\bibitem [{\citenamefont {Haken}(2004)}]{Haken2004}%
  \BibitemOpen
  \bibfield  {author} {\bibinfo {author} {\bibfnamefont {H.}~\bibnamefont
  {Haken}},\ }\bibinfo {title} {An introduction},\ in\ \href
  {https://doi.org/10.1007/978-3-662-10184-1_1} {\emph {\bibinfo {booktitle}
  {Synergetics: Introduction and Advanced Topics}}}\ (\bibinfo  {publisher}
  {Springer Berlin Heidelberg},\ \bibinfo {address} {Berlin, Heidelberg},\
  \bibinfo {year} {2004})\ pp.\ \bibinfo {pages} {1--387}\BibitemShut {NoStop}%
\bibitem [{\citenamefont {Haken}(1977)}]{Haken_1977}%
  \BibitemOpen
  \bibfield  {author} {\bibinfo {author} {\bibfnamefont {H.}~\bibnamefont
  {Haken}},\ }\bibfield  {title} {\bibinfo {title} {Synergetics},\ }\href
  {https://doi.org/10.1088/0031-9112/28/9/027} {\bibfield  {journal} {\bibinfo
  {journal} {Phys. Bull.}\ }\textbf {\bibinfo {volume} {28}},\ \bibinfo {pages}
  {412} (\bibinfo {year} {1977})}\BibitemShut {NoStop}%
\bibitem [{\citenamefont {Haake}(1982)}]{haake1982systematic}%
  \BibitemOpen
  \bibfield  {author} {\bibinfo {author} {\bibfnamefont {F.}~\bibnamefont
  {Haake}},\ }\bibfield  {title} {\bibinfo {title} {Systematic adiabatic
  elimination for stochastic processes},\ }\href@noop {} {\bibfield  {journal}
  {\bibinfo  {journal} {Z. Phys. B - Condensed Matter}\ }\textbf {\bibinfo
  {volume} {48}},\ \bibinfo {pages} {31} (\bibinfo {year} {1982})}\BibitemShut
  {NoStop}%
\bibitem [{\citenamefont {Sancho}\ \emph {et~al.}(1982)\citenamefont {Sancho},
  \citenamefont {Miguel},\ and\ \citenamefont
  {D{\"u}rr}}]{sancho1982adiabatic}%
  \BibitemOpen
  \bibfield  {author} {\bibinfo {author} {\bibfnamefont {J.}~\bibnamefont
  {Sancho}}, \bibinfo {author} {\bibfnamefont {M.~S.}\ \bibnamefont {Miguel}},\
  and\ \bibinfo {author} {\bibfnamefont {D.}~\bibnamefont {D{\"u}rr}},\
  }\bibfield  {title} {\bibinfo {title} {Adiabatic elimination for systems of
  brownian particles with nonconstant damping coefficients},\ }\href@noop {}
  {\bibfield  {journal} {\bibinfo  {journal} {J. Stat. Phys.}\ }\textbf
  {\bibinfo {volume} {28}},\ \bibinfo {pages} {291} (\bibinfo {year}
  {1982})}\BibitemShut {NoStop}%
\bibitem [{\citenamefont {Gardiner}(1984)}]{gardiner1984adiabatic}%
  \BibitemOpen
  \bibfield  {author} {\bibinfo {author} {\bibfnamefont {C.}~\bibnamefont
  {Gardiner}},\ }\bibfield  {title} {\bibinfo {title} {Adiabatic elimination in
  stochastic systems. i. formulation of methods and application to few-variable
  systems},\ }\href@noop {} {\bibfield  {journal} {\bibinfo  {journal} {Phys.
  Rev. A}\ }\textbf {\bibinfo {volume} {29}},\ \bibinfo {pages} {2814}
  (\bibinfo {year} {1984})}\BibitemShut {NoStop}%
\bibitem [{\citenamefont {Lugiato}\ \emph {et~al.}(1984)\citenamefont
  {Lugiato}, \citenamefont {Mandel},\ and\ \citenamefont
  {Narducci}}]{lugiato1984adiabatic}%
  \BibitemOpen
  \bibfield  {author} {\bibinfo {author} {\bibfnamefont {L.~A.}\ \bibnamefont
  {Lugiato}}, \bibinfo {author} {\bibfnamefont {P.}~\bibnamefont {Mandel}},\
  and\ \bibinfo {author} {\bibfnamefont {L.}~\bibnamefont {Narducci}},\
  }\bibfield  {title} {\bibinfo {title} {Adiabatic elimination in nonlinear
  dynamical systems},\ }\href@noop {} {\bibfield  {journal} {\bibinfo
  {journal} {Phys. Rev. A}\ }\textbf {\bibinfo {volume} {29}},\ \bibinfo
  {pages} {1438} (\bibinfo {year} {1984})}\BibitemShut {NoStop}%
\bibitem [{\citenamefont {Gough}\ and\ \citenamefont
  {Van~Handel}(2007)}]{gough2007singular}%
  \BibitemOpen
  \bibfield  {author} {\bibinfo {author} {\bibfnamefont {J.}~\bibnamefont
  {Gough}}\ and\ \bibinfo {author} {\bibfnamefont {R.}~\bibnamefont
  {Van~Handel}},\ }\bibfield  {title} {\bibinfo {title} {Singular perturbation
  of quantum stochastic differential equations with coupling through an
  oscillator mode},\ }\href@noop {} {\bibfield  {journal} {\bibinfo  {journal}
  {J. Stat. Phys.}\ }\textbf {\bibinfo {volume} {127}},\ \bibinfo {pages} {575}
  (\bibinfo {year} {2007})}\BibitemShut {NoStop}%
\bibitem [{\citenamefont {Lax}(1966)}]{lax1966quantum}%
  \BibitemOpen
  \bibfield  {author} {\bibinfo {author} {\bibfnamefont {M.}~\bibnamefont
  {Lax}},\ }\bibfield  {title} {\bibinfo {title} {Quantum noise. iv. quantum
  theory of noise sources},\ }\href@noop {} {\bibfield  {journal} {\bibinfo
  {journal} {Phys. Rev.}\ }\textbf {\bibinfo {volume} {145}},\ \bibinfo {pages}
  {110} (\bibinfo {year} {1966})}\BibitemShut {NoStop}%
\bibitem [{\citenamefont {Wiseman}\ and\ \citenamefont
  {Milburn}(2009)}]{wiseman2009quantum}%
  \BibitemOpen
  \bibfield  {author} {\bibinfo {author} {\bibfnamefont {H.~M.}\ \bibnamefont
  {Wiseman}}\ and\ \bibinfo {author} {\bibfnamefont {G.~J.}\ \bibnamefont
  {Milburn}},\ }\href@noop {} {\emph {\bibinfo {title} {Quantum measurement and
  control}}}\ (\bibinfo  {publisher} {Cambridge university press},\ \bibinfo
  {year} {2009})\BibitemShut {NoStop}%
\bibitem [{\citenamefont {Gardiner}\ and\ \citenamefont
  {Zoller}(2004)}]{gardiner2004quantum}%
  \BibitemOpen
  \bibfield  {author} {\bibinfo {author} {\bibfnamefont {C.}~\bibnamefont
  {Gardiner}}\ and\ \bibinfo {author} {\bibfnamefont {P.}~\bibnamefont
  {Zoller}},\ }\href@noop {} {\emph {\bibinfo {title} {Quantum noise: a
  handbook of Markovian and non-Markovian quantum stochastic methods with
  applications to quantum optics}}}\ (\bibinfo  {publisher} {Springer Science
  \& Business Media},\ \bibinfo {year} {2004})\BibitemShut {NoStop}%
\bibitem [{\citenamefont {Duan}\ \emph {et~al.}(2001)\citenamefont {Duan},
  \citenamefont {Lukin}, \citenamefont {Cirac},\ and\ \citenamefont
  {Zoller}}]{duan2001long}%
  \BibitemOpen
  \bibfield  {author} {\bibinfo {author} {\bibfnamefont {L.-M.}\ \bibnamefont
  {Duan}}, \bibinfo {author} {\bibfnamefont {M.~D.}\ \bibnamefont {Lukin}},
  \bibinfo {author} {\bibfnamefont {J.~I.}\ \bibnamefont {Cirac}},\ and\
  \bibinfo {author} {\bibfnamefont {P.}~\bibnamefont {Zoller}},\ }\bibfield
  {title} {\bibinfo {title} {Long-distance quantum communication with atomic
  ensembles and linear optics},\ }\href@noop {} {\bibfield  {journal} {\bibinfo
   {journal} {Nature}\ }\textbf {\bibinfo {volume} {414}},\ \bibinfo {pages}
  {413} (\bibinfo {year} {2001})}\BibitemShut {NoStop}%
\bibitem [{\citenamefont {Sangouard}\ \emph {et~al.}(2011)\citenamefont
  {Sangouard}, \citenamefont {Simon}, \citenamefont {de~Riedmatten},\ and\
  \citenamefont {Gisin}}]{RevModPhys.83.33}%
  \BibitemOpen
  \bibfield  {author} {\bibinfo {author} {\bibfnamefont {N.}~\bibnamefont
  {Sangouard}}, \bibinfo {author} {\bibfnamefont {C.}~\bibnamefont {Simon}},
  \bibinfo {author} {\bibfnamefont {H.}~\bibnamefont {de~Riedmatten}},\ and\
  \bibinfo {author} {\bibfnamefont {N.}~\bibnamefont {Gisin}},\ }\bibfield
  {title} {\bibinfo {title} {Quantum repeaters based on atomic ensembles and
  linear optics},\ }\href {https://doi.org/10.1103/RevModPhys.83.33} {\bibfield
   {journal} {\bibinfo  {journal} {Rev. Mod. Phys.}\ }\textbf {\bibinfo
  {volume} {83}},\ \bibinfo {pages} {33} (\bibinfo {year} {2011})}\BibitemShut
  {NoStop}%
\bibitem [{\citenamefont {S\o{}rensen}\ and\ \citenamefont
  {M\o{}lmer}(2002)}]{PhysRevA.66.022314}%
  \BibitemOpen
  \bibfield  {author} {\bibinfo {author} {\bibfnamefont {A.~S.}\ \bibnamefont
  {S\o{}rensen}}\ and\ \bibinfo {author} {\bibfnamefont {K.}~\bibnamefont
  {M\o{}lmer}},\ }\bibfield  {title} {\bibinfo {title} {Entangling atoms in bad
  cavities},\ }\href {https://doi.org/10.1103/PhysRevA.66.022314} {\bibfield
  {journal} {\bibinfo  {journal} {Phys. Rev. A}\ }\textbf {\bibinfo {volume}
  {66}},\ \bibinfo {pages} {022314} (\bibinfo {year} {2002})}\BibitemShut
  {NoStop}%
\bibitem [{\citenamefont {Doherty}\ \emph {et~al.}(1998)\citenamefont
  {Doherty}, \citenamefont {Parkins}, \citenamefont {Tan},\ and\ \citenamefont
  {Walls}}]{doherty1998motional}%
  \BibitemOpen
  \bibfield  {author} {\bibinfo {author} {\bibfnamefont {A.}~\bibnamefont
  {Doherty}}, \bibinfo {author} {\bibfnamefont {A.}~\bibnamefont {Parkins}},
  \bibinfo {author} {\bibfnamefont {S.}~\bibnamefont {Tan}},\ and\ \bibinfo
  {author} {\bibfnamefont {D.}~\bibnamefont {Walls}},\ }\bibfield  {title}
  {\bibinfo {title} {Motional states of atoms in cavity qed},\ }\href@noop {}
  {\bibfield  {journal} {\bibinfo  {journal} {Phys. Rev. A}\ }\textbf {\bibinfo
  {volume} {57}},\ \bibinfo {pages} {4804} (\bibinfo {year}
  {1998})}\BibitemShut {NoStop}%
\bibitem [{\citenamefont {Duan}\ and\ \citenamefont
  {Kimble}(2004)}]{duan2004scalable}%
  \BibitemOpen
  \bibfield  {author} {\bibinfo {author} {\bibfnamefont {L.-M.}\ \bibnamefont
  {Duan}}\ and\ \bibinfo {author} {\bibfnamefont {H.}~\bibnamefont {Kimble}},\
  }\bibfield  {title} {\bibinfo {title} {Scalable photonic quantum computation
  through cavity-assisted interactions},\ }\href@noop {} {\bibfield  {journal}
  {\bibinfo  {journal} {Phys. Rev. Lett.}\ }\textbf {\bibinfo {volume} {92}},\
  \bibinfo {pages} {127902} (\bibinfo {year} {2004})}\BibitemShut {NoStop}%
\bibitem [{\citenamefont {Hofer}\ \emph {et~al.}(2011)\citenamefont {Hofer},
  \citenamefont {Wieczorek}, \citenamefont {Aspelmeyer},\ and\ \citenamefont
  {Hammerer}}]{PhysRevA.84.052327}%
  \BibitemOpen
  \bibfield  {author} {\bibinfo {author} {\bibfnamefont {S.~G.}\ \bibnamefont
  {Hofer}}, \bibinfo {author} {\bibfnamefont {W.}~\bibnamefont {Wieczorek}},
  \bibinfo {author} {\bibfnamefont {M.}~\bibnamefont {Aspelmeyer}},\ and\
  \bibinfo {author} {\bibfnamefont {K.}~\bibnamefont {Hammerer}},\ }\bibfield
  {title} {\bibinfo {title} {Quantum entanglement and teleportation in pulsed
  cavity optomechanics},\ }\href {https://doi.org/10.1103/PhysRevA.84.052327}
  {\bibfield  {journal} {\bibinfo  {journal} {Phys. Rev. A}\ }\textbf {\bibinfo
  {volume} {84}},\ \bibinfo {pages} {052327} (\bibinfo {year}
  {2011})}\BibitemShut {NoStop}%
\bibitem [{\citenamefont {Galland}\ \emph {et~al.}(2014)\citenamefont
  {Galland}, \citenamefont {Sangouard}, \citenamefont {Piro}, \citenamefont
  {Gisin},\ and\ \citenamefont {Kippenberg}}]{PhysRevLett.112.143602}%
  \BibitemOpen
  \bibfield  {author} {\bibinfo {author} {\bibfnamefont {C.}~\bibnamefont
  {Galland}}, \bibinfo {author} {\bibfnamefont {N.}~\bibnamefont {Sangouard}},
  \bibinfo {author} {\bibfnamefont {N.}~\bibnamefont {Piro}}, \bibinfo {author}
  {\bibfnamefont {N.}~\bibnamefont {Gisin}},\ and\ \bibinfo {author}
  {\bibfnamefont {T.~J.}\ \bibnamefont {Kippenberg}},\ }\bibfield  {title}
  {\bibinfo {title} {Heralded single-phonon preparation, storage, and readout
  in cavity optomechanics},\ }\href
  {https://doi.org/10.1103/PhysRevLett.112.143602} {\bibfield  {journal}
  {\bibinfo  {journal} {Phys. Rev. Lett.}\ }\textbf {\bibinfo {volume} {112}},\
  \bibinfo {pages} {143602} (\bibinfo {year} {2014})}\BibitemShut {NoStop}%
\bibitem [{\citenamefont {Gerry}\ and\ \citenamefont
  {Knight}(2023)}]{gerry2023introductory}%
  \BibitemOpen
  \bibfield  {author} {\bibinfo {author} {\bibfnamefont {C.~C.}\ \bibnamefont
  {Gerry}}\ and\ \bibinfo {author} {\bibfnamefont {P.~L.}\ \bibnamefont
  {Knight}},\ }\href@noop {} {\emph {\bibinfo {title} {Introductory quantum
  optics}}}\ (\bibinfo  {publisher} {Cambridge university press},\ \bibinfo
  {year} {2023})\BibitemShut {NoStop}%
\bibitem [{\citenamefont {Aspelmeyer}\ \emph {et~al.}(2014)\citenamefont
  {Aspelmeyer}, \citenamefont {Kippenberg},\ and\ \citenamefont
  {Marquardt}}]{RevModPhys.86.1391}%
  \BibitemOpen
  \bibfield  {author} {\bibinfo {author} {\bibfnamefont {M.}~\bibnamefont
  {Aspelmeyer}}, \bibinfo {author} {\bibfnamefont {T.~J.}\ \bibnamefont
  {Kippenberg}},\ and\ \bibinfo {author} {\bibfnamefont {F.}~\bibnamefont
  {Marquardt}},\ }\bibfield  {title} {\bibinfo {title} {Cavity optomechanics},\
  }\href {https://doi.org/10.1103/RevModPhys.86.1391} {\bibfield  {journal}
  {\bibinfo  {journal} {Rev. Mod. Phys.}\ }\textbf {\bibinfo {volume} {86}},\
  \bibinfo {pages} {1391} (\bibinfo {year} {2014})}\BibitemShut {NoStop}%
\bibitem [{\citenamefont {Collett}\ and\ \citenamefont
  {Gardiner}(1984)}]{collett1984squeezing}%
  \BibitemOpen
  \bibfield  {author} {\bibinfo {author} {\bibfnamefont {M.}~\bibnamefont
  {Collett}}\ and\ \bibinfo {author} {\bibfnamefont {C.}~\bibnamefont
  {Gardiner}},\ }\bibfield  {title} {\bibinfo {title} {Squeezing of intracavity
  and traveling-wave light fields produced in parametric amplification},\
  }\href@noop {} {\bibfield  {journal} {\bibinfo  {journal} {Phys. Rev. A}\
  }\textbf {\bibinfo {volume} {30}},\ \bibinfo {pages} {1386} (\bibinfo {year}
  {1984})}\BibitemShut {NoStop}%
\bibitem [{\citenamefont {Walls}\ and\ \citenamefont
  {Milburn}(2008)}]{walls2008input}%
  \BibitemOpen
  \bibfield  {author} {\bibinfo {author} {\bibfnamefont {D.}~\bibnamefont
  {Walls}}\ and\ \bibinfo {author} {\bibfnamefont {G.~J.}\ \bibnamefont
  {Milburn}},\ }\href@noop {} {\emph {\bibinfo {title} {Quantum optics}}}\
  (\bibinfo  {publisher} {Springer},\ \bibinfo {year} {2008})\BibitemShut
  {NoStop}%
\bibitem [{\citenamefont {Stein}\ and\ \citenamefont
  {Shakarchi}(2010)}]{stein2010complex}%
  \BibitemOpen
  \bibfield  {author} {\bibinfo {author} {\bibfnamefont {E.~M.}\ \bibnamefont
  {Stein}}\ and\ \bibinfo {author} {\bibfnamefont {R.}~\bibnamefont
  {Shakarchi}},\ }\href@noop {} {\emph {\bibinfo {title} {Complex analysis}}},\
  Vol.~\bibinfo {volume} {2}\ (\bibinfo  {publisher} {Princeton University
  Press},\ \bibinfo {year} {2010})\BibitemShut {NoStop}%
\bibitem [{\citenamefont {Lindell}(1993)}]{lindell1993delta}%
  \BibitemOpen
  \bibfield  {author} {\bibinfo {author} {\bibfnamefont {I.~V.}\ \bibnamefont
  {Lindell}},\ }\bibfield  {title} {\bibinfo {title} {Delta function
  expansions, complex delta functions and the steepest descent method},\
  }\href@noop {} {\bibfield  {journal} {\bibinfo  {journal} {Am. J. Phys.}\
  }\textbf {\bibinfo {volume} {61}},\ \bibinfo {pages} {438} (\bibinfo {year}
  {1993})}\BibitemShut {NoStop}%
\end{thebibliography}%

\end{document}